\documentclass{article}

\PassOptionsToPackage{numbers, compress}{natbib}

     \usepackage[final]{neurips_2022}

\usepackage[utf8]{inputenc} 
\usepackage[T1]{fontenc}    
\usepackage{hyperref}       
\usepackage{url}            
\usepackage{booktabs}       
\usepackage{amsfonts}       
\usepackage{nicefrac}       
\usepackage{microtype}      
\usepackage{xcolor}         

\title{An Analytics of Culture: Modeling Subjectivity, Scalability, Contextuality, and Temporality}

\author{%
Nanne van Noord\thanks{Joint first authorship. Order of first authors may swap depending on context.} \\
Informatics Institute\\
University of Amsterdam \\
\texttt{n.j.e.vannoord@uva.nl}\\
 \And
Melvin Wevers\footnotemark[1] \\
Department of History\\
University of Amsterdam\\
\texttt{m.j.h.f.wevers@uva.nl} \\
 \And
Tobias Blanke \\
Institute for Logic, Language and Computation \\
University of Amsterdam\\
\texttt{t.blanke@uva.nl} \\
 \And
Julia Noordegraaf \\
Department of Media Studies \\
University of Amsterdam \\
\texttt{j.j.noordegraaf@uva.nl} \\
 \And
Marcel Worring \\
Informatics Institute \\
University of Amsterdam \\
\texttt{m.worring@uva.nl}
}

\begin{document}

\maketitle

\section{Introduction}

There is a bidirectional relationship between culture and AI; AI models are increasingly used to analyse culture, thereby shaping our understanding of culture. On the other hand, the models are trained on collections of cultural artifacts thereby implicitly, and not always correctly, encoding expressions of culture. This creates a tension that both limits the use of AI for analysing culture and leads to problems in AI with respect to cultural complex issues such as bias. 

One approach to overcome this tension is to more extensively take into account the intricacies and complexities of culture. We structure our discussion using four concepts that guide humanistic inquiry into culture: subjectivity, scalability, contextuality, and temporality. We focus on these concepts because they have not yet been sufficiently represented in AI research. We believe that possible implementations of these aspects into AI research leads to AI that better captures the complexities of culture. 
In what follows, we briefly describe these four concepts and their absence in AI research. For each concept, we define possible research challenges. 

\section{Subjectivity} 
In analysing culture, we often want to incorporate and compare viewpoints by different subjects, providing us with a more holistic perspective on culture. Subjectivity is inherently relational, meaning that viewpoints gain importance through their relationship with other viewpoints. These subjectivities are not represented in a monolithic model that is a simple amalgamation of viewpoints. Consequently, it is difficult to identify and contrast viewpoints in a model. 

Frequency-based AI approaches may not always accurately represent the prevalence of viewpoints in society, leading to an over-representation of minority and/or undesirable viewpoints. Attempts have been made to de-bias models by removing undesired viewpoints or to emphasise desirable viewpoints~\cite{bolukbasiManComputerProgrammer2016, weversUsingWordEmbeddings2019} 
Rather than simply removing diversity, the research challenge is to explicitly encode viewpoints into a model, or into multiple models. The result could be models representing specific personas; from elites to underrepresented groups. For example, when training an embedding model on historical textual data, one could encapsulate social class into the model. 

\section{Scalability} 
Cultural analysis understands cultural phenomena as inherently multi-scale through time and space, working though the interactions of particularity and universality. Digital cultural data can be viewed at different levels, allowing for `close' and `distant reading/viewing'. However, this binary categorization overlooks the interactions between the cultural phenomena on different levels. We need AI systems to be able to scale down to the micro-levels of individual interactions and up to the macro-levels of collective transformation across varying modalities~\cite{carrignonTablewareTradeRoman2020, leeScalingTheoryArmedconflict2020}. 

When modelling cultural data for AI research, we have to take into consideration these interactions between scales and the commonalities across them. Otherwise, we run the risk of conflating culture into one big melting pot, overlooking differences between individuals and groups as well as continuities across groups. We need theoretical frameworks to explain large-scale trends through small-scale processes. This also means that we have to carefully assess which parts of humanistic inquiry can be automated and scaled using AI systems and which parts require human interpretation or human-machine interaction.

\section{Contextuality} 
Cultural artifacts obtain their meaning and significance from the context in which they originate, circulate, and are perceived. 
Traditionally, a humanities scholar studies the cultural artifact and its context and learns to intuit what is meaningful and what is not, thereby advancing their understanding and insights about the artifact. In addition to the artifact itself and its metadata, therefore, a fuller understanding of its cultural value requires data that represents these contexts. In a review of context, \citet{Bradley_Dunlop_2005} make the distinction between incidental and meaningful context. Where the incidental context is that which just happens to be present, whereas meaningful context may influence the process of extracting information for obtaining insight. The challenge when analysing culture is to determine which contextual information is incidental and which is meaningful. 

Working with archival data requires knowledge and expertise to adequately determine what is meaningful. Information offered through metadata can only in part assist in determining what is meaningful. To be able to make the distinguishing between incidental and meaningful, we need data that goes beyond the \textit{where}, \textit{when}, and \textit{by whom} an artifact originated, but also includes the \textit{why}~\cite{Noordegraaf_2015}. Currently, many models ignore the metadata outright and deal with the artifact data only. The open question is how can computational models be derived that not only have sufficient `base’ knowledge to contextualise but that are also able to distinguish between the incidental and meaningful? Additionally, how can we take into account that not all meaningful context might be available as data? Situated Data Analysis might offer entry points for this latter question \cite{Rettberg_2020}.

\section{Temporality} 
Time is crucial for cultural analysis in part because the order of events influences meaning formation~\cite{weversEventFlowHow2021}. The meaning of concepts relating to sexual identity, such as queer, or the visual representation of objects, such as telephones, cannot be accurately modelled without considering temporal changes. Currently, in analysing cultural data with AI this temporal dimension is ignored, thereby disregarding processes related to intertextuality, intericonicity, and the formation of canons \cite{vanNoord_2022}. A cultural expression's ability to refer to earlier expressions, or its position in a canon determines how we view them and what role they play. 

Datasets are often temporal snapshots, capturing a single moment in time and the meaning that concepts had at that point~\cite{Blanke_2018}. Currently, having multiple snapshots allows us to contrast and compare different senses, but it does not allow us to capture their meaning evolved. To be able to capture how meaning is constituted through cultural processes, AI models need to encode how meaning forms over time and how meanings are retained and possibly forgotten over time~\cite{tassierModelFadsFashions2004}. Relying on longitudinal data by itself does not offer the solution, the challenge is in finding the relationality between temporal data points and the principles driving how cultural data is produced over time.

\section{Conclusion}

We have described four concepts of the humanistic analysis of culture and the challenges they present for AI. We contend that there is a fruitful trading zone where humanistic and AI scholarship can meet and shape the development of AI. As such, this paper provides an invitation for productive engagement and collaboration between domains to do justice to the complexity of human culture.

\bibliography{blib}

\begin{thebibliography}{11}
\providecommand{\natexlab}[1]{#1}
\providecommand{\url}[1]{\texttt{#1}}
\expandafter\ifx\csname urlstyle\endcsname\relax
  \providecommand{\doi}[1]{doi: #1}\else
  \providecommand{\doi}{doi: \begingroup \urlstyle{rm}\Url}\fi

\bibitem[Blanke(2018)]{Blanke_2018}
Tobias Blanke.
\newblock Predicting the past.
\newblock \emph{Digital Humanities Quarterly}, 12\penalty0 (2):\penalty0
  1–21, Jul 2018.
\newblock ISSN 1938-4122.

\bibitem[Bolukbasi et~al.()Bolukbasi, Chang, Zou, Saligrama, and
  Kalai]{bolukbasiManComputerProgrammer2016}
Tolga Bolukbasi, Kai-Wei Chang, James~Y Zou, Venkatesh Saligrama, and Adam~T
  Kalai.
\newblock Man is to {{Computer Programmer}} as {{Woman}} is to {{Homemaker}}?
  {{Debiasing Word Embeddings}}.
\newblock In D.~D. Lee, M.~Sugiyama, U.~V. Luxburg, I.~Guyon, and R.~Garnett,
  editors, \emph{Advances in {{Neural Information Processing Systems}} 29},
  pages 4349--4357. {Curran Associates, Inc.}

\bibitem[Bradley and Dunlop(2005)]{Bradley_Dunlop_2005}
Nicholas~A Bradley and Mark~D Dunlop.
\newblock Towards a multidisciplinary model of context to support context-aware
  computing.
\newblock \emph{Journal of Human-Computer Interaction}, 20\penalty0
  (4):\penalty0 403–446, 2005.

\bibitem[Carrignon et~al.()Carrignon, Brughmans, and
  Romanowska]{carrignonTablewareTradeRoman2020}
Simon Carrignon, Tom Brughmans, and Iza Romanowska.
\newblock Tableware trade in the {{Roman East}}: {{Exploring}} cultural and
  economic transmission with agent-based modelling and approximate {{Bayesian}}
  computation.
\newblock 15\penalty0 (11):\penalty0 e0240414.

\bibitem[Lee et~al.()Lee, Daniels, Myers, Krakauer, and
  Flack]{leeScalingTheoryArmedconflict2020}
Edward~D. Lee, Bryan~C. Daniels, Christopher~R. Myers, David~C. Krakauer, and
  Jessica~C. Flack.
\newblock Scaling theory of armed-conflict avalanches.
\newblock 102\penalty0 (4):\penalty0 042312.

\bibitem[Noordegraaf(2015)]{Noordegraaf_2015}
Julia Noordegraaf.
\newblock Crowdsourcing television’s past: The state of knowledge in digital
  archives.
\newblock \emph{TMG Journal for Media History}, 14\penalty0 (22):\penalty0
  108–120, Sep 2015.
\newblock ISSN 2213-7653.
\newblock \doi{10.18146/tmg.139}.

\bibitem[Rettberg(2020)]{Rettberg_2020}
Jill~Walker Rettberg.
\newblock Situated data analysis: a new method for analysing encoded power
  relationships in social media platforms and apps.
\newblock \emph{Humanities and Social Sciences Communications}, 7\penalty0
  (11):\penalty0 1–13, Jun 2020.
\newblock ISSN 2662-9992.
\newblock \doi{10.1057/s41599-020-0495-3}.

\bibitem[Tassier()]{tassierModelFadsFashions2004}
Troy Tassier.
\newblock A model of fads, fashions, and group formation.
\newblock 9\penalty0 (5):\penalty0 51--61.

\bibitem[van Noord(2022)]{vanNoord_2022}
Nanne van Noord.
\newblock A survey of computational methods for iconic image analysis.
\newblock \emph{Digital Scholarship in the Humanities}, 2022.
\newblock ISSN 2055-7671.
\newblock \doi{10.1093/llc/fqac003}.

\bibitem[Wevers()]{weversUsingWordEmbeddings2019}
Melvin Wevers.
\newblock Using {{Word Embeddings}} to {{Examine Gender Bias}} in {{Dutch
  Newspapers}}, 1950-1990.
\newblock In \emph{Proceedings of the 1st {{International Workshop}} on
  {{Computational Approaches}} to {{Historical Language Change}}}, pages
  92--97. {Association for Computational Linguistics}.

\bibitem[Wevers et~al.()Wevers, Kostkan, and Nielbo]{weversEventFlowHow2021}
Melvin Wevers, Jan Kostkan, and Kristoffer~L. Nielbo.
\newblock Event {{Flow}} - {{How Events Shaped}} the {{Flow}} of the {{News}},
  1950-1995.
\newblock In Maud Ehrmann, Folgert Karsdorp, Melvin Wevers, Tara~Lee Andrews,
  Manuel Burghardt, Mike Kestemont, Enrique Manjavacas, Michael Piotrowski, and
  Joris van Zundert, editors, \emph{Proceedings of the {{Conference}} on
  {{Computational Humanities}} {{Research}} 2021}, volume 2989 of \emph{{{CEUR
  Workshop Proceedings}}}, pages 62--76. {CEUR}.

\end{thebibliography}
\bibliographystyle{plainnat}

\end{document}